\documentclass[12pt]{article}
\usepackage{latexsym,amsfonts,amsmath,amsthm,amssymb}
\begin{document}

\title{Quantum-like Representation of Extensive Form Games: Wine Testing Game}
\author{Andrei Khrennikov\\
International Center for
Mathematical Modeling \\ in Physics and Cognitive Sciences\\
University of V\"axj\"o, S-35195, Sweden}

\maketitle

\begin{abstract} We consider an application of the mathematical formalism of quantum mechanics (QM)
outside physics, namely, to game theory. We present a simple game between macroscopic players, say
Alice and Bob (or in a more complex form - Alice, Bob and Cecilia), which can be represented in the
quantum-like (QL) way -- by using a complex probability amplitude (game's ``wave function'') and noncommutative
operators. The crucial point is that games under consideration are so called extensive form games. Here 
the order of actions of players is important, such a game can be represented by the tree of actions.
The QL probabilistic behavior of players is a consequence of incomplete information which is available to 
e.g. Bob about the previous action of Alice. In general one could not construct a classical probability space
underlying a QL-game. This can happen even in a QL-game with two players. In a QL-game with 
three players Bell's inequality can be violated. The most natural probabilistic
 description is given by so called contextual probability 
theory completed by the frequency definition of probability.
\end{abstract}

\section{Introduction}

One should sharply distinguish between QM as a physical theory and the mathematical formalism of QM. In the same way 
as one should distinguish between classical Newtonian mechanics and its mathematical formalism. Nobody is surprised
that differential and integral calculi which are basic in Newtonian mechanics can be fruitfully applied in other 
domains of science. Unfortunately, the situation  with the mathematical formalism of QM is essentially more complicated --
some purely mathematical specialities of QM are projected on and even identified with specialities 
of quantum physical systems. Although already Nils Bohr pointed out by himself \cite{Bohr}, see also \cite{PL1},
\cite{PL2},  to the possibility 
to apply the mathematical formalism of QM outside of physics, the prejudices based on identification of 
mathematics and physics still survive (but cf. e.g. Accardi, Ballentine, De Muynck, Gudder, Land\'e, Mackey 
\cite{AC1}--\cite{MC1} and also \cite{FPP}--\cite{FPP3}) 
and one can point out just to a few applications outside of physics. Here we discuss not {\it reductionist models} in that the quantum description appears as a consequence 
of composing of a system (for example, the brain, see e.g. \cite{P1}, \cite{P2}) of quantum particles, 
but really the possibility to use the mathematical 
formalism of QM without direct coupling with quantum physics, see e.g. \cite{KH1}, \cite{KH2}.     

One of interesting possibilities to explore quantum mathematics is provided by game theory \cite{VN1}, \cite{OW}. 
One of the main 
distinguishing features  of the mathematical formalism of QM is the calculus of quantum probabilities. 
It is the calculus \cite{D}, \cite{VN} of complex probability amplitudes and self-adjoint operators,
in contrast to the calculus of random variables on the  Kolmogorov classical
probability space \cite{KOL}. The impossibility to use a fixed Kolmogorov probability space induces 
applications of such a probabilistic theory as {\it Gudder's theory of probabilistic manifolds}, see also Accardi 
\cite{AC2} for non-Kolmogorovian models. Recently Karl Hess and Walter Philipp pointed out to the old paper of Soviet 
mathematician Vorobjev \cite{VR} who studied the problem of the possibility to realize a number of observables
on a single Kolmogorv space. This problem is equivalent to the problem of violation of Bell's inequality which
was later studied by J. Bell \cite{BEL}. However, nonlocality was not involved in Vorobjev's considerations.
It is especially interesting for us that Vorobjev pointed out to a possibility 
to apply probabilistic models without underlaying Kolmogorov structure in game theory (in fact, he promised to come with 
such applications in later publications, but I was able not find anything). One may consider the present paper 
as a step toward realization of the Vorobjev's program. 

We present a simple game between macroscopic players, say
Alice and Bob (or in a more complex form - Alice, Bob and Cecilia), which can be represented in the
quantum-like (QL) way -- by using a complex probability amplitude (game's ``wave function'') and noncommutative
operators. The crucial point is that  games under consideration are so called extensive form games, see e.g.
\cite{EF}. Here 
the order of actions of players is important, such a game can be represented by the tree of actions.
The QL probabilistic behavior of players is a consequence of incomplete information which is available to 
e.g. Bob about the previous action of Alice. In general one could not construct a classical probability space
underlying a QL-game. This can happen even in a QL-game with two players. In a QL-game with 
three players Bell's inequality can be violated. The most 
natural probabilistic description is given by so called contextual probability 
theory \cite{KHC} completed by the frequency definition of probability \cite{MI}, \cite{KH3}. 
In particular this theory provides 
an algorithm -- {\it quantum-like representation algorithm} (QLRA) \cite{KHC} 
which gives a possibility to represent 
special collections of probabilistic data for a pair of observables $a,b$ by complex probability amplitudes
(or in the abstract formalism by normalized vectors of the complex Hilbert space) and observables by self-adjoint 
operators $\hat{a}, \hat{b}.$ We shall use QLRA to find QL-representations of extensive form games.
The probabilistic structure of our game (for two players) can be considered as Gudder's probability 
manifold \cite{GD0} with the atlas having two charts.

The first examples of QL-games with macroscopic players were presented in papers \cite{GR1}--\cite{GR3} which were based essentially
on quantum logic models \cite{GR4}, \cite{GR5}. In this paper we use contextual probabilistic arguments which provide
a possibility 
to take into a more detailed account the probabilistic sources of  the QL-behaviour of players. 

Our examples  are totally
different from what now is widely discussed in many papers in the
name of quantum games \cite{Ek}. All examples with quantum
coins, quantum gamblers etc. in this or that way use micro objects
described by quantum physics as some hardware, while in our
examples everything is totally macroscopic. However, some results obtained 
in the cited quantum game activity can be applied to our examples 
(because we use the same mathematical apparatus).

Our study of QL-games can also be considered as a contribution in clarification of problems in foundations of QM. 
In particular, the problems of ``death of reality'' and nonlocality, cf. \cite{ES}, \cite{Sh}. Analysis of QL-games supports the Bohr's viewpoint 
on observables -- the result of a measurement cannot be considered as an objective property of this 
systems which could be assigned to the system before the measurement starts. In our models this results are preferences 
of Alice and Bob in choice of wine as well as their abilities to test wine. Such things could not be assigned
with a bottle of wine as its objective properties. Nevertheless, game theory shows that there are no reasons for 
panics -- death of reality. There is reality of wine and its chemical properties as well as reality of brains which 
induces finally results of measurements. 

The QL-behavior can be produced in the purely local framework. However, a game can be completed by  
interactions between players (of course, laws of special relativity are not violated). Such games are even more interesting and they could have more extended domain of 
applications.     

\section{Contextual probability}
 
A general statistical model for observables based on the contextual
viewpoint to probability will be presented. It will be shown that
classical as well as quantum probabilistic models can be obtained as
particular cases of our general contextual model, the {\it{V\"axj\"o
model}}, \cite{KHC}.
    
This model is not reduced to the conventional, classical and quantum
    models. In particular, it contains a new statistical model: a model with
hyperbolic $cosh$-interference that induces  "hyperbolic quantum
mechanics" \cite{KHC}.

A physical, biological, social,  mental, genetic, economic, or financial
{\it context}  $C$ is  a complex of corresponding conditions.
Contexts are fundamental elements of any contextual statistical model. Thus construction  
of any model
$M$ should be started with fixing the collection of  contexts of this model.
Denote the collection of contexts
by the symbol ${\cal C}$ (so the family of contexts  ${\cal C}$ is determined by the 
model $M$ under consideration). In the mathematical formalism ${\cal C}$ is an abstract set
(of ``labels'' of contexts).  

We remark that in some models it is possible to construct a set-theoretic 
representation of contexts -- as some family of subsets of a set $\Omega.$ For example,
$\Omega$ can be the set of all possible parameters (e.g., physical, or mental, or economic)
of the model. However, in general we {\it do not assume the possibility to construct a set-theoretic 
representation of contexts.}

Another fundamental element of any contextual statistical model 
$M$ is a set of observables ${\cal O}:$
each observable $a\in {\cal O}$ can be measured
under each complex of conditions $C\in {\cal C}.$  
For an observable $a \in {\cal O},$ we denote the set
of its possible values (``spectrum'') by the symbol
$X_a.$

We do not assume that all these observables can  be measured
simultaneously. To simplify considerations, we shall consider only
discrete observables and, moreover, all concrete investigations will
be performed for {\it dichotomous observables.}

\medskip

{\bf Axiom 1:} {\it For  any observable
$a \in {\cal O}$  and its value $\alpha \in X_a,$ there are defined contexts, say $C_\alpha,$
corresponding to $\alpha$-selections: if we perform a measurement of the observable $a$ under
the complex of physical conditions $C_\alpha,$ then we obtain the value $a=\alpha$ with
probability 1. We assume  that the set of contexts ${\cal C}$ contains 
$C_\alpha$-selection contexts for all observables $a\in {\cal O}$ and $\alpha \in X_a.$}

\medskip

For example, let $a$ be the observable corresponding to some question:  $a=+$ (the answer ``yes'')
and $a=-$ (the answer ``no''). Then the $C_{+}$-selection context is the selection of 
those participants of the experiment who answering ``yes'' 
to this question; in the same way we define the  
$C_{-}$-selection context. By Axiom 1 these contexts are well defined. 
We point out that in principle a participant of this experiment might not want to reply at all 
to this question or she  might change her mind immediately after her answer.
By Axiom 1 such possibilities are excluded. By the same 
axiom both $C_{+}$ and $C_{-}$-contexts belong to the system of contexts under consideration.

\medskip

{\bf Axiom 2:} {\it There are defined contextual (conditional) 
probabilities $p_C^a(\alpha) \equiv {\bf P}(a= \alpha\vert C)$ for any
context $C \in {\cal C}$ and any observable $a \in {\it O}.$}

\medskip

Thus, for any context $C \in {\cal C}$ and any observable $a \in {\it O},$ 
there is defined the probability to observe the fixed value $a=\alpha$ under the complex 
of conditions $C.$

Especially important role will be played by ``transition probabilities''
$
p^{a\vert b}(\alpha\vert \beta)\equiv {\bf P}(a=\alpha\vert C_\beta), a, b \in {\cal O}, \alpha \in X_a, \beta \in X_b,
$
where $C_\beta$ is the $[b=\beta]$-selection context. By axiom 2 for any context $C\in {\cal C},$ 
there is defined the set of probabilities:
$
 \{ p_C^a : a \in {\cal O}\}.
$
We complete this probabilistic data for the context $C$  by transition probabilities.
  The corresponding collection of data $D({\cal O}, C)$ 
consists of contextual probabilities:
$
p^{a\vert b}(\alpha\vert \beta),p_C^b(\beta),
p^{b\vert a}(\beta \vert \alpha), p_C^a(\alpha)...,
$
where $a,b,... \in {\cal O}.$ Finally, we denote  the family of
probabilistic data $D({\cal O}, C)$ for all contexts  $C\in {\cal
C}$ by the symbol ${\cal D}({\cal O}, {\cal C}) 
(\equiv \cup_{C\in {\cal C}} D({\cal O}, C)).$

\medskip

{\bf Definition 1.} (V\"axj\"o Model) {\it An observational contextual  statistical model of reality is a triple
$M =({\cal C}, {\cal O}, {\cal D}({\cal O}, {\cal C})),$
where ${\cal C}$ is a set of contexts and ${\cal O}$ is a  set of observables
which satisfy to axioms 1,2, and ${\cal D}({\cal O}, {\cal C})$ is probabilistic data
about contexts ${\cal C}$ obtained with the aid of observables belonging ${\cal O}.$}

\medskip

We call observables belonging the set ${\cal O}\equiv {\cal O}(M)$ {\it reference of observables.}
Inside of a model $M$  observables  belonging  to the set ${\cal O}$ give the only possible references
about a context $C\in {\cal C}.$ In the definition of the V\"axj\"o Model we speak about ``reality.''
In our approach it is reality of contexts. 

In what follows we shall consider V\"axj\"o models with two dichotomous reference observables.

\section{Frequency definition of probabilities}

The definition of probability has not yet been specified. In this paper we shall use the frequency definition of probability 
as the limit of frequencies in a long series of trials, von Mises' approach, \cite{MI}, , \cite{KH3}. We are aware that this approach 
was criticized a lot in mathematical literature. However, the main critique was directed against von Mises' 
definition of randomness. If one is not interested in randomness, but only in frequencies of trials, then the 
frequency approach is well established, see  \cite{KH3}. 

We consider a set of reference
observables ${\cal O}= \{ a, b \}$ consisting of two observables $a$ and $b.$
We denotes the sets of values (``spectra'') of the reference observables by symbols $X_a$ and $X_b,$
respectively.

Let $C$ be some context. In a series of observations of $b$ (which can be infinite in a mathematical model)
we obtain a sequence of values of $b:$
$x\equiv x(b \vert C) = (x_1, x_2,..., x_N,...), \;\; x_j\in X_b.$
In a series of observations of $a$ we obtain a sequence of values of $a:$
$y\equiv y(a \vert C) = (y_1, y_2,..., y_N,...), \;\; y_j\in X_a.$
We suppose that the {\it principle of the statistical stabilization} for relative frequencies \cite{MI}, \cite{KH3}
holds. This means that the frequency probabilities
are well defined:
$p_C^b(\beta) = \lim_{N\to \infty} \nu_N(\beta; x), \;\; \beta \in X_b;$
$p_C^a(\alpha)= \lim_{N\to \infty} \nu_N(\alpha; y), \;\; \alpha\in X_a.$
Here $\nu_N(\beta; x)$ and $ \nu_N(\alpha; y)$ are frequencies of observations of values
$b=\beta$ and $a=\alpha,$ respectively (under the complex of conditions $C).$

{\bf Remark.} (On the notions of collective and $S$-sequence) R. von Mises considered
in his theory two principles: a) the principle of the statistical stabilization for relative frequencies;
 b) the principle of randomness. A sequence of observations for which
both principle hold was called a {\it collective,} \cite{MI}. 
However, it seems that the validity of the principle
of statistical stabilization  is often enough  for applications. Here we shall use just the
convergence of frequencies to probabilities. An analog of von Mises' theory for sequences of observations
 which satisfy the principle of statistical stabilization
was developed in \cite{KH3}; we call such sequences $S$-{\it sequences.}

Everywhere in this paper it will be assumed that {\it sequences of observations are $S$-sequences},
cf. \cite{KH3} (so we are not interested in the validity of the principle of randomness for sequences
of observations, but only in existence of the limits of relative frequencies).

Let $C_{\alpha},  \alpha\in X_a,$  be contexts  corresponding
to  $\alpha$-filtrations, see Axiom 1.
By observation of $b$ under the context $C_\alpha$ we obtain a sequence:
$x^{\alpha} \equiv x(b \vert C_\alpha) = (x_1, x_2,..., x_{N},...), \;\; x_j \in X_b.$
It is also assumed that for  sequences of observations  $x^{\alpha}, \alpha\in X_a,$
the principle of statistical stabilization for relative frequencies
holds true and the frequency probabilities are well defined:
$p^{b \vert a}(\beta \vert \alpha) = \lim_{N \to \infty} \nu_{N}(\beta; x^{\alpha}), \;\;
\beta \in X_b.$
Here $\nu_N(\beta; x^\alpha), \alpha\in X_a,$  are frequencies of observations of value
$b=\beta$ under the complex of conditions $C_\alpha.$
We can repeat all previous considerations by changing $b\vert a$-conditioning to  $a \vert b$-conditioning.
There can be defined probabilities $p^{a \vert b}(\alpha \vert \beta).$

\section{Quantum-like representation algorithm -- QLRA}

In \cite{KHC} we derived the following formula for interference of probabilities:
\begin{equation}
\label{TFR} p_C^b(\beta) = \sum_\alpha p_C^a(\alpha) p^{b\vert
a}(\beta\vert \alpha) + 2 \lambda(\beta\vert  \alpha, C)
\sqrt{\prod_\alpha p_C^a(\alpha) p^{b\vert a}(\beta\vert \alpha)},
\end{equation}
where the coefficient of interference 
\begin{equation}
\label{KOL6}
\lambda(\beta\vert  a, C) = \frac{p_C^b(\beta) - \sum_\alpha p_C^a(\alpha) p^{b\vert
a}(\beta\vert \alpha)}{2 \sqrt{\prod_\alpha p_C^a(\alpha) p^{b\vert a}(\beta\vert \alpha)}} .
\end{equation}
A similar representation we have for the $a$-probabilities.
Such interference formulas are valid for any collection of contextual 
probabilistic data satisfying the conditions:

\medskip

R1). Observables $a$ and $b$ are symmetrically conditioned:
$$
p^{b\vert a}( \beta \vert  \alpha)= p^{a \vert b}( \alpha \vert  \beta).
$$ 

\medskip

R2). $p^{a\vert b}( \alpha \vert  \beta)>0$ and $p^{b \vert a}( \beta\vert  \alpha)>0$ as well
as   $p_C^b(\beta) >0$ and $p_C^a(\alpha) >0.$

\medskip

Suppose that also the following conditions hold:

\medskip

R3). Coefficients of interference $\lambda( \beta\vert a, C)$ and $\lambda( \alpha\vert b, C)$ are bounded by one.
 
\medskip

A context $C$ such that R3) holds is called trigonometric, because in this case we have the conventional formula of 
trigonometric interference: 
\begin{equation}
\label{TNCZ} p_C^b(\beta) = \sum_\alpha p_C^a(\alpha) p^{b\vert
a}(\beta\vert \alpha) + 2 \cos\theta(\beta\vert  \alpha, C)
\sqrt{\prod_\alpha p_C^a(\alpha) p^{b\vert a}(\beta\vert \alpha)},
\end{equation}
where $
\lambda(\beta\vert a ,C)=\cos \theta (\beta\vert a,C).
$
Parameters $\theta(\beta\vert \alpha,C)$ are said to be $b \vert
a$-{\it relative phases} with respect to the context $C.$ We defined these phases 
purely on the basis of probabilities. We have not started with any linear space; in contrast we shall
define geometry from probability.

We denote the collection of all trigonometric contexts by the symbol ${\cal C}^{\rm{tr}}.$

By using the elementary formula:
$$
D=A+B+2\sqrt{AB}\cos \theta=\vert \sqrt{A}+e^{i \theta}\sqrt{B}|^2,
$$
for real numbers $A, B > 0, \theta\in [0,2 \pi],$ we can represent
the probability $p_C^b(\beta)$ as the square of the complex
amplitude (Born's rule):
\begin{equation}
\label{Born} p_C^b(\beta)=\vert \psi_C(\beta) \vert^2 \;.
\end{equation}
Here
\begin{equation}
\label{EX1} \psi(\beta) \equiv \psi_C(\beta)=
\sqrt{p_C^a(\alpha_1)p^{b\vert a}(\beta\vert \alpha_1)} + e^{i
\theta_C(\beta)} \sqrt{p_C^a(\alpha_2)p^{b\vert a}(\beta\vert
\alpha_2)}, \; \beta \in X_b,
\end{equation}
where $\theta_C(\beta)\equiv \theta(\beta\vert  \alpha, C).$

\medskip

The formula (\ref{EX1}) gives the quantum-like representation
algorithm -- QLRA. For any trigonometric context $C$ by starting
with the probabilistic data -- $ p_C^b(\beta), p_C^a(\alpha),
p^{b\vert a}(\beta\vert \alpha)$ -- QLRA produces the complex
amplitude $ \psi_C.$ This algorithm can be used in any domain of
science to create the QL-representation of probabilistic data (for a
special class of contexts).

We point out that QLRA contains the reference observables as
parameters. Hence the complex amplitude give by (\ref{EX1}) depends
on $a,b: \psi_C \equiv \psi_C^{b\vert a}.$

We denote the space of functions: $\varphi: X_b\to {\bf C}$ by the
symbol $\Phi =\Phi(X_b, {\bf C}).$ Since $X= \{\beta_1, \beta_2 \},$ the
$\Phi$ is the two dimensional complex linear space. By using QLRA
 we construct the map $J^{b \vert a}:{\cal C}^{\rm{tr}}
\to \Phi(X, {\bf C})$ which maps contexts (complexes of, e.g.,
physical conditions) into complex amplitudes. The representation
({\ref{Born}}) of probability is nothing other than the famous {\bf
Born rule.} The complex amplitude $\psi_C(x)$ can be called a
{\bf wave function} of the complex of physical conditions (context)
$C$  or a  (pure) {\it state.}  We set $e_\beta^b(\cdot)=\delta(\beta-
\cdot)$ -- Dirac delta-functions concentrated in points $\beta=
\beta_1, \beta_2.$ The Born's rule for complex amplitudes (\ref{Born}) can be
rewritten in the following form: $\label{BH}
p_C^b(\beta)=\vert \langle \psi_C, e_\beta^b \rangle \vert^2,$ where the scalar product
in the space $\Phi(X_b, C)$ is defined by the standard formula:
$\langle \phi, \psi \rangle = \sum_{\beta\in X_b} \phi(\beta)\bar \psi(\beta).$ The system
of functions $\{e_\beta^b\}_{\beta\in X_b}$ is an orthonormal basis in the
Hilbert space $H_{ab}=(\Phi, \langle \cdot, \cdot \rangle).$ 

Let $X_b \subset {\bf R}.$ By using
the Hilbert space representation  of the Born's rule  we
obtain  the Hilbert space representation of the expectation of the
observable $b$: $E(b \vert C)= \sum_{\beta\in
X_b} \beta\vert\psi_C(\beta)\vert^2= \sum_{\beta\in X_b} \beta \langle \psi_C, e_\beta^b\rangle
\overline{\langle\psi_C, e_\beta^b\rangle}= \langle \hat b \psi_C, \psi_C\rangle,$ where
the  (self-adjoint) operator $\hat b: H_{ab} \to H_{ab}$ is determined by its
eigenvectors: $\hat b e_\beta^b=\beta e^b_\beta, \beta\in X_b.$ This is the
multiplication operator in the space of complex functions
$\Phi(X_b,{\bf C}):$ $ \hat{b} \psi(\beta) = \beta \psi(\beta).$ It is
natural to represent the $b$-observable  (in the Hilbert space
model)  by the operator $\hat b.$ 

We would like to have Born's rule
not only for the $b$-variable, but also for the $a$-variable:
$p_C^a(\alpha)=\vert \langle \varphi, e_\alpha^a \rangle\vert^2 \;, \alpha \in  X_a.$

How can we define the basis $\{e_\alpha^a\}$ corresponding to the
$a$-observable? Such a basis can be found starting with interference
of probabilities. We set $u_j^a=\sqrt{p_C^a(\alpha_j)},
p_{ij}=p(\beta_j \vert \alpha_i), u_{ij}=\sqrt{p_{ij}}, \theta_j=\theta_C(\beta_j).$ We
have:
\begin{equation}
\label{0} \varphi=u_1^a e_{\alpha_1}^a + u_2^a e_{\alpha_2}^a,
\end{equation}
where
\begin{equation}
\label{Bas} e_{\alpha_1}^a= (u_{11}, \; \; u_{12}) ,\; \; e_{\alpha_2}^a= (e^{i
\theta_1} u_{21}, \; \; e^{i \theta_2} u_{22})
\end{equation}
The condition R1) implies that the  system $\{e_{\alpha_i}^a\}$  is an orthonormal basis iff 
the probabilistic phases satisfy the
constraint:
$$
\theta_2 - \theta_1= \pi \; \rm{mod} \; 2 \pi,
$$  but, as we have seen \cite{KHC}, we can always choose such  phases (under the condition R1).

In this case the $a$-observable is represented by the operator
$\hat{a}$ which is diagonal with eigenvalues $\alpha_1,\alpha_2$ in the basis
$\{e_{\alpha_i}^a\}.$ The  conditional average of the observable
 $a$ coincides with the quantum Hilbert space average:
$
E(a \vert C)=\sum_{\alpha \in X_a} \alpha p_C^a(\alpha) = \langle \hat{a} \psi_C, \psi_C \rangle.
$

\section{Wine testing game}

There is restaurant having a good collection of (only) French and Italian wines of various sorts. 
Couples come to this restaurant for dinners and to have more fun they play the following Wine Game
which consists of two wine tests. 

\medskip

A1). Alice selects  a bottle (without to tell her friend Bob wine's name) 
and proposes him to test wine. A battle of this wine is opened in restaurant's kitchen, Bob gets just a glass of this wine. 
Alice asks him the question:

\medskip

{\it ``Is it  French or Italian?''}

\medskip
\medskip

A3). If Bob answers (after testing) correctly, he gets some amount of money; if not, he looses money 
and Alice gets some amount of money.

\medskip
The choice in A1 is not totally  random, Alice has her own preferences (later she wants to share the chosen bottle with Bob). 

In the second part of the game Alice and Bob interchange their roles, so Bob starts by choosing a bottle 
of  French or Italian 
wines and so on.

We introduce for the first and second parts of the game the 
elements of the payment matrices 
$$
(h_{FF;k}^b,h_{FI;k}^b,...), \;(h_{FF;k}^a, h_{FI;k}^a,...), k=1,2.
$$ 
Here the indexes $k =1,2$ denote the first and second part of the game and FI,..., II combinations 
of choices of Alice and Bob.\footnote{We also remark the Alice's choice can be considered as an ``element of reality'', since her, e.g.,  F,
is really French wine, but Bob's F may be in reality either French or Italian wine, cf. with discussions about realism 
in quantum mechanics, e.g., \cite{ES}, \cite{Sh}.} 
 The upper indexes $a,b$ are marks for Alice's and Bob's payoffs. It is natural to assume that 
$$
h_{FF;1}^b, h_{II;1}^b >0, \; \; h_{FI;1}^b, h_{IF;1}^b < 0
$$ 
as well as 
$$
h_{FI;1}^a, h_{FI;1}^a >0,\; \;  h_{FF;1}^a, h_{II;1}^a < 0.
$$ 
In the zero sum game  
$$
h_{FF;k}^b= - h_{FF;k}^a,\; ..., \; h_{II;k}^b= - h_{II;k}^a.
$$ 

Each part of this game can be represented as an {\it extensive form game}, hence, by a tree, see \cite{OS}, \cite{EF}
(or just the link http://en.wikipedia.org). 
This tree is very simple and it has the following branches representing actions 
of Alice and Bob; each branch is finished by the pair of payoffs, 
the symbol "v" used for vertexes and "act" for 
corresponding actions. The first part of the game is represented by the tree with the branches:

\medskip

$v=A- act=F-v=B-act=F-(h_{FF;1}^b,h_{FF;1}^a);$

$v=A-act=F-v=B-act=I-h_{FI;1}^b, h_{FI;1}^b);$ 

$v=A-act=I-v=B-act=F-(h_{IF;1}^b,h_{IF;1}^a);$  

$v=A-act=I-v=B-act=I-(h_{II;1}^b,h_{II;1}^a).$  

\medskip

As always, we are interested in averages of wins-losses of Alice and Bob.

We consider the following probabilities:

\medskip

1). Probabilities of Alice's preferences for a bottle of French wine and respectivly a bottle of Italian wine from 
the wine-collection of the restaurant: 
$$
p_C^a(F), \; p_C^a(I).
$$ 
Here the index $C$ is related to the whole context of the game, in particular, to the
collection of wines. Another restaurant has another collection of wines, and 
Alice would have other preferences.

2). Probabilities to recognize French wine after testing (by Bob) 
a bottle of French wine which was chosen  by Alice for the test:
$p^{b\vert a}(F\vert F);$ the probability of mistake under this 
condition, i.e., claiming that the wine is Italian, 
is then $p^{b\vert a}(I\vert F)= 1- p^{b\vert a}(F\vert F).$ In the similar way we introduce probabilities
$p^{b\vert a}(I\vert I)$  and $p^{b\vert a}(F\vert I).$ Thus we have the matrix which is typically called 
the {\it matrix of transition probabilities}:
$$
{\bf P}^{b \vert a}=(p^{b\vert a}( \beta\vert \alpha )), \;  \beta, \alpha =I, F.
$$

3). Similarly  we introduce probabilities $p_C^b(F)$ and $p_C^b(I)$ for Bob's preferences 
(for the same collection of wines) as well as probabilities
$p^{a \vert b}(F\vert F), ..., \\p^{a \vert b}(I \vert I)$ which represent Alice's ability to recognize 
the origin of wine. There is the matrix of transition probabilities 
${\bf P}^{a \vert b}=(p^{a\vert b}( \alpha \vert  \beta)), \; \alpha ,  \beta=I, F. $

4). Finally, we introduce probabilities that Bob will announce the result $\beta (=F,I)$ in the game that 
Alice starts with the result $\alpha$ (which is hidden from Bob): 
$$
p_C^{ab}(\alpha, \beta)= p_C^a(\alpha) p^{b\vert a}(\beta\vert \alpha),
$$ and similar 
probabilities for the game which is started by Bob: $p_C^{ba}(\beta, \alpha).$

We remark that $p_C^{ab}(\alpha, \beta)$ is really probability on the set of all pairs $(\alpha, \beta):$
$$
\sum_{\alpha,\beta}  p_C^{ab}(\alpha, \beta)=\sum_{\alpha} p_C^a(\alpha) \sum_{\beta} 
p^{b\vert a}(\beta\vert \alpha)=
1.
$$
This probability serves well for the first part of the game -- when Alice chooses a bottle:
$$
p_C^a(\alpha)=\sum_{\beta}  p_C^{ab}(\alpha, \beta).
$$
However, it could not be used in the second part of the game, since in general:
$$
p_C^b(\beta)\not=\sum_{\alpha}  p_C^{ab}(\alpha, \beta).
$$
The second part of the game is served by the probability $p_C^{ba}(\beta, \alpha).$
The tricky thing, see \cite{GR3}, is really the combination of two games.

\medskip

We point out that in general the equality  
\begin{equation}
\label{BFH}
p_C^{ab}(\alpha, \beta) = p_C^{ba}(\beta, \alpha)
\end{equation}
can be violated. This is the main source of ``nonclassicality'' of our game.

\medskip
Then the average wins-losses in the first 
part of the game  for Bob is given by
$$
E_1^b(C)= 
h_{FF;1}^b \; p_C^{ab}(F,F) + h_{FI;1}^b \;p_C^{ab}(F,I) 
+ h_{IF;1}^b \; p_C^{ab}(I,F) + h_{II;1}^b \;p_C^{ab}(I,I).
$$  
The average for Alice in the first part of the game  (in general we can consider nonzero sum game) is given by
$$
E_1^a(C)= h_{FF;1}^a \; p_C^{ab}(F,F) + h_{FI;1}^a \;p_C^{ab}(F,I)
+ h_{IF;1}^a \;p_C^{ab}(I,F) + h_{II;1}^a \;p_C^{ab}(I,I).
$$
In the same way the averages for Alice and Bob in the second part of the game are given by 
$$
E_2^a(C)= h_{FF;2}^a \;p_C^{ba}(F,F) + h_{FI;2}^a \;p_C^{ba}(F,I)  +
h_{IF;2}^a \;p_C^{ba}(I,F) + h_{II;2}^a \;p_C^{ba}(I, I).
$$
$$
E_2^b(C)= h_{FF;2}^b \; p_C^{ba}(F,F)  + h_{FI;2}^b \;p_C^{ba}(F,I) +
h_{IF;2}^b \;p_C^{ba}(I,F) + h_{II;2}^b \;p_C^{ba}(I, I).
$$
The averages of total wins-losses are 
$$
E^b(C)= E_1^b(C) + E_2^b(C), \; E^a(C)= E_1^a(C) + E_2^a(C).
$$ 
It is convenient to introduce a ``wine-observable'' for Alice:
$a=F,I.$  This observable appears in two different contexts.
The first context, $C,$ is the context of selection of a bottle from the wine 
collection. Alice  chooses a bottle and says herself (not Bob!)
or just think -- it is French wine (or it is Italian wine). The second context appears in the second 
part of the game when Alice should test wine proposed by Bob and after that say: it is French wine (or it is Italian wine).
In fact, to be completely correct one should consider two different observables corresponding to these contexts. However,
to have closer analogy with quantum mechanics, we proceed with one observable. Alice is considered as simply an apparatus
which says either ``French wine'' or ``Italian wine'' (cf. with Stern-Gerlach magnet, it ``says'' either ``spin up''
or ``spin down''). We remark that our cognitive example shows that it might be more natural to associate
with each quantum state -- wave function -- its own spin-observable.
We introduce a similar observable for Bob, $b= F,I.$

\section{Extensive form game with imperfect information}

As was mentioned, formally wine testing game is an extensive form game. However, we should point out to one rather delicate
feature of the game. We recall that a complete extensive form representation specifies:
   1) the players of a game;
   2) for every player every opportunity they have to move;
   3) what each player can do at each of their moves;
   4) what each player knows for every move;
   5) the payoffs received by every player for every possible combination of moves.

Our game fulfills all those conditions besides the fourth one. In fact, the action of Alice does not specify
for Bob the result of her action, Bob should guess about the country origin of the wine given by Alice. To come to 
the conclusion, he should perform a rather complicated analysis of the wine test. One may say that this is a game with 
{\it imperfect information.} 

We recall that an {\it information set} is a set of decision nodes such that:
1) every node in the set belongs to one player;
2) when play reaches the information set, the player with the move cannot differentiate between nodes 
within the information set, i.e. if the information set contains more than one node, 
the player to whom that set belongs does not know which node in the set has been reached.

If a game has an information set with more than one member that game is said to have imperfect information. 
A game with perfect information is such that at any stage of the game, every player knows exactly 
what has taken place earlier in the game, i.e. every information set is a singleton set. 
Any game without perfect information has imperfect information. 

However, there is a problem with the second condition determining the information set. Of course, Bob does not know precisely 
which kind of wine is presented for the test. In this sense the set of Bob's nodes after Alice's action (we consider the 
first part of the game) forms an information set. But (and this is crucial) Bob has the possibility to analyze wine 
(cf. with measurement process in quantum physics). Therefore he might distinguish two actions of Alice, F and I, but only 
{\it partially.} I have no idea whether such a problem of analysis of actions of the opposite player was discussed in game theory?

\section{Quantum-like representation}

The wine testing game has  a natural QL-representation. 
Let us consider a game with restrictions R1)--R3) on strategies, see
section 4.
By applying QLRA to statistical data we can construct 
a probability amplitude $\psi_C(\beta), \beta=F, I.$ 
To simplify considerations, we assume that the coefficients of  interference
are bounded by one. Thus the context $C$ is trigonometric and the 
probability amplitude is complex valued. It can also be represented by a unit vector 
of the two dimensional complex Hilbert space. We remark that in principle there are no reasons 
for such an assumption. In opposite to QM, Wine Game might produce hyperbolic probability 
amplitudes \cite{KHC}.

 In this case we can represent the wins-losses
averages in the QL-way: 
$$
E^b(C)=h_{FF;1}^b \; \vert \langle \psi_C, e_F^a \rangle\vert^2 \;\vert \langle e_F^b, e_F^a \rangle \vert^2 
 + h_{IF;1}^b \;\vert \langle \psi_C, e_I^a \rangle\vert^2 \;\vert \langle e_F^b, e_I^a \rangle \vert^2
$$ 
$$
+ h_{FI;1}^b \;\vert \langle \psi_C, e_F^a \rangle \vert^2 \;\vert \langle e_I^b, e_F^a \rangle \vert^2
+ h_{II;1}^b \; \vert \langle \psi_C, e_I^a \rangle \vert^2 \;\vert \langle e_I^b, e_I^a \rangle \vert^2
$$
$$
+ h_{FF;2}^b \; \vert \langle \psi_C, e_F^b \rangle\vert^2 \;\vert \langle e_F^b, e_F^a \rangle \vert^2
 + h_{IF;2}^b \;\vert \langle \psi_C, e_I^b \rangle\vert^2 \;\vert \langle e_I^b, e_F^a \rangle \vert^2
$$ 
$$
+ h_{FI;2}^b \;\vert \langle \psi_C, e_F^b \rangle\vert^2 \;\vert \langle e_F^b, e_I^a \rangle \vert^2
+ h_{II;2}^b \;\vert \langle \psi_C, e_I^b \rangle\vert^2 \;\vert \langle e_I^b, e_I^a \rangle \vert^2 .
$$
In the same way we represent the average for Alice.
Thus the wine testing game satisfying conditions R1-R3 can be represented in the complex Hilbert space.

The QL-expression for the average is essentially simpler in the case of zero sum game with symmetry between the first and second
parts: $h_{FF;1}^b=h_{FF;2}^a= -h_{FF;2}^b, ..., h_{II;1}^b=h_{II;2}^a= - h_{II;2}^b.$ Here
$$
E^b(C) =h_{FF;1}^b (\vert \langle \psi_C, e_F^a \rangle\vert^2 \;\vert \langle e_F^b, e_F^a \rangle \vert^2 -
\vert \langle \psi_C, e_F^b \rangle\vert^2 \;\vert \langle e_F^b, e_F^a \rangle \vert^2)$$
$$
 + h_{IF;1}^b (\vert \langle \psi_C, e_I^a \rangle\vert^2 \;\vert \langle e_F^b, e_I^a \rangle \vert^2 -
 \vert \langle \psi_C, e_I^b \rangle\vert^2 \;\vert \langle e_I^b, e_F^a \rangle \vert^2)
$$ 
$$
+ h_{FI;1}^b (\vert \langle \psi_C, e_F^a \rangle \vert^2 \;\vert \langle e_I^b, e_F^a \rangle \vert^2 -
\vert \langle \psi_C, e_F^b \rangle\vert^2 \;\vert \langle e_F^b, e_I^a \rangle \vert^2)
$$
$$
+ h_{II;1}^b \; (\vert \langle \psi_C, e_I^a \rangle \vert^2 \;\vert \langle e_I^b, e_I^a \rangle \vert^2-
\vert \langle \psi_C, e_I^b \rangle\vert^2 \;\vert \langle e_I^b, e_I^a \rangle \vert^2)
$$
$$
=h_{FF;1}^b \;\vert \langle e_F^b, e_F^a \rangle \vert^2\; (\vert \langle \psi_C, e_F^a \rangle\vert^2  -
\vert \langle \psi_C, e_F^b \rangle\vert^2)
 + h_{IF;1}^b \;\vert \langle e_F^b, e_I^a \rangle \vert^2\; (\vert \langle \psi_C, e_I^a \rangle\vert^2  -
 \vert \langle \psi_C, e_I^b \rangle\vert^2)
$$ 
$$
+ h_{FI;1}^b \;\vert \langle e_I^b, e_F^a \rangle \vert^2\; (\vert \langle \psi_C, e_F^a \rangle \vert^2  -
\vert \langle \psi_C, e_F^b \rangle\vert^2)
+ h_{II;1}^b \; \vert \langle e_I^b, e_I^a \rangle \vert^2\; (\vert \langle \psi_C, e_I^a \rangle \vert^2 -
\vert \langle \psi_C, e_I^b \rangle\vert^2)
$$
$$
=(\vert \langle \psi_C, e_F^a \rangle\vert^2  -
\vert \langle \psi_C, e_F^b \rangle\vert^2)\; (h_{FF;1}^b \;\vert \langle e_F^b, e_F^a \rangle \vert^2\; 
 + h_{FI;1}^b \;\vert \langle e_I^b, e_F^a \rangle \vert^2)
$$ 
$$
+ (\vert \langle \psi_C, e_I^a \rangle \vert^2  -
\vert \langle \psi_C, e_I^b \rangle\vert^2)\; (h_{IF;1}^b \;\vert \langle e_F^b, e_I^a \rangle \vert^2\; 
+ h_{II;1}^b \; \vert \langle e_I^b, e_I^a \rangle \vert^2).
$$

\section{Superposition of preferences}

We now point out that we can expand e.g. vectors of the $b$-basis with respect to the $a$-basis:
$e_F^b= c_{FF} e_F^a + c_{FI} e_I^a, e_I^b= c_{IF} e_F^a + c_{II} e_I^a.$ One might say that ``Bob's preferences 
are superpositions of Alice preferences.'' However, we cannot assign any real meaning to such a sentence
in  the present game framework. Thus superposition is merely a purely mathematical representation -- the geometric
picture of the probabilistic structure of the game. In the same way we can expand the state $\psi_C$ 
with respect to the $a$-basis as well as the $b$-basis. Such expansions neither have any real meaning, just 
geometrical representation of probabilities. Nevertheless, such a picture is convenient for geometric representation
of mental states of Alice and Bob. One may use the following geometric model: there are two basic 
mental states of Alice (in the context of Wine Game) $e_F^a$ and $e_I^a.$ In general Alice plays in the 
superposition of these states $\psi=c_F^a e_F^a + c_I^a e_I^a.$ In the same way Bob has two basic mental states
$e_F^b$ and $e_I^b.$  In general Bob plays in the 
superposition of these states $\psi=c_F^b e_F^b + c_I^b e_I^b.$ Moreover, (at least mathematically) Bob's mental states
can be represented as superpositions of Alice's mental states.

We can represent the average of Bob's wins-losses in the interference form:   
$$
E^b(C)=(\vert \langle \psi_C, e_F^a \rangle\vert^2  -
\vert \bar{c}_{FF} \langle \psi_C, e_F^a\rangle + \bar{c}_{FI}
 \langle \psi_C, e_I^a\rangle \vert^2\;)\; (h_{FF;1}^b \;\vert \langle e_F^b, e_F^a \rangle \vert^2\; 
 + h_{FI;1}^b \;\vert \langle e_I^b, e_F^a \rangle \vert^2)
$$ 
$$
+ (\vert \langle \psi_C, e_I^a \rangle \vert^2  -
\vert  \bar{c}_{IF} \langle \psi_C, e_F^a\rangle + \bar{c}_{II} \langle \psi_C, e_I^a\rangle \vert^2\;)\; (h_{IF;1}^b \;\vert \langle e_F^b, e_I^a \rangle \vert^2\; 
+ h_{II;1}^b \; \vert \langle e_I^b, e_I^a \rangle \vert^2)
$$
$$
=(\vert \langle \psi_C, e_F^a \rangle\vert^2  
-(\vert \langle \psi_C, e_F^a\rangle\vert^2 \vert \langle e_F^b, e_F^a\rangle\vert^2 +
\vert \langle \psi_C, e_I^a\rangle\vert^2 \vert \langle e_F^b, e_I^a\rangle\vert^2 
$$
$$
+2\cos\theta
\vert \langle \psi_C, e_F^a\rangle 
\langle e_F^b, e_F^a\rangle \langle \psi_C, e_I^a\rangle \langle e_F^b, e_I^a\rangle\vert))\;
 (h_{FF;1}^b \;\vert \langle e_F^b, e_F^a \rangle \vert^2\; 
 + h_{FI;1}^b \;\vert \langle e_I^b, e_F^a \rangle \vert^2)
$$ 
$$
+ (\vert \langle \psi_C, e_I^a \rangle \vert^2  -
(\vert \langle \psi_C, e_F^a\rangle\vert^2 \vert \langle e_I^b, e_F^a\rangle\vert^2 +
\vert \langle \psi_C, e_I^a\rangle\vert^2 \vert \langle e_I^b, e_I^a\rangle\vert^2 
$$
$$
- 2\cos\theta
\vert \langle \psi_C, e_F^a\rangle 
\langle e_I^b, e_F^a\rangle \langle \psi_C, e_I^a\rangle \langle e_I^b, e_I^a\rangle\vert)\;)
\; (h_{IF;1}^b \;\vert \langle e_F^b, e_I^a \rangle \vert^2\; 
+ h_{II;1}^b \; \vert \langle e_I^b, e_I^a \rangle \vert^2).
$$

\section{Meaning of the wave function}

The wave function $\psi_C$ was constructed on the basis of probabilities:   
$p_C^a( \alpha), p_C^a(\beta), \\ p^{b\vert a}(\beta \vert  \alpha).$ It represents the wine collection of the restaurant 
as well as preferences of  Alice and Bob. Moreover, it also represents 
their abilities  to find difference between French and Italian wines. 

Thus one may say such a wave function (complex probability amplitude) 
$\psi_C$ has no real counterpart. We could not point out to any object in reality 
which is represented by $\psi_C.$ It represents
the context of the wine collection as well as Bob's and Alice's preferences and 
experiences with different kinds of wines.
Such a context  is extremely complex. It is impossible to describe its precisely. However, the $\psi_C$ 
provides some approximative representation of this context in the complex Hilbert space.

We notify that Alice and Bob are coupled through the wave function. The wave function really provides a possibility
to combine probabilistic features of two cognitive systems, Alice and Bob, which could not be incorporated
into a single Kolmogorov probability space.

\section{The role of Bayes formula}

Suppose at the moment that randomness of actions of Alice and Bob can be described by the Kolmogorov 
probability space ${\cal P}=(\Omega, {\cal F}, {\bf P})$ in the following way:

a). The wine-collection context $C$ is represented by an element of ${\cal F}$ which will be denoted 
by the same symbol.

b). Probabilities 
$$
p_C^a(\alpha)={\bf P }_C (A_\alpha) \equiv  \frac{{\bf P}(A_\alpha \cap C)}{{\bf P}(C)}, 
p_C^b(\beta) ={\bf P }_C (B_\beta) \equiv \frac{{\bf P}(B_\beta \cap C)}{{\bf P}(C)},
$$ 
where 
$$
A_\alpha=\{ \omega \in \Omega:
a(\omega)=\alpha \}, \; B_\beta =\{ \omega \in \Omega:
b(\omega)=\beta \}.
$$
Here ${\bf P}_C$ is the conditional probability measure corresponding to the subset $C$ of 
${\cal F}: {\bf P}_C(A)= \frac{{\bf P}(A\cap C)}{{\bf P}(C)}.$

c). Transition portabilities 
$$
p^{b \vert a}(\beta \vert \alpha) = {\bf P }_C (B_\beta \vert A_\alpha) \equiv
\frac{{\bf P}(B_\beta \cap A_\alpha \cap C)}{{\bf P}(A_\alpha \cap C)}.
$$

\medskip

In such a representation the $C$-conditional Bayes formula holds:
\begin{equation}
\label{BF}
{\bf P}_C (A_\alpha \cap B_\beta) = {\bf P}_C (A_\alpha) {\bf P}_C (B_\beta \vert A_\alpha).
\end{equation}  
Hence, here the equality (\ref{BFH}) holds! (Because the Kolmogorovian probability is symmetric:
${\bf P}_C (A_\alpha \cap B_\beta)= {\bf P}_C (B_\beta\cap A_\alpha).)$ We obtain the following equality, see
\cite{KHC} for details:   
\begin{equation}
\label{BF1}
{\bf P}_C (A_\alpha) {\bf P}_C (B_\beta \vert A_\alpha)=
{\bf P}_C (B_\beta) {\bf P}_C ( A_\alpha \vert B_\beta).
\end{equation}
Since we want to get the QL-representation, we consider symmetrically conditioned variables $a$ and $b,$ see R1).
The condition (\ref{BF1}) implies that 
\begin{equation}
\label{BF2}
{\bf P}_C(A_\alpha)= {\bf P}_C(B_\beta) = 1/2.
\end{equation}
Thus one can construct a Kolmogorov representation of Wine Game satisfying conditions a)-c) iff
selection of wines from collection is uniformly distributed between French and Italian wines 
(both for Alice and Bob). If not, then {\it there is no Kolmogorov model.} For example,
if  the probability that Alice chooses
a bottle of French wine $p_C^a(F)=1/3$ (and consequently the probability 
that she chooses a bottle of Italian wine $p_C^a(I)=2/3),$ then it is impossible to construct a Kolmogorov 
probability space for this game. Of course, one should not forget that we assumed that the game probabilities
are coupled to the Kolmogorov space via conditions a)-c) and that we would like to have symmetric transition
 probabilities.

The origin of this nonclassicality of the probabilistic description is {\it impossibility to combine on a single 
Kolmogorov space preferences of Alice and Bob in choosing wines and their abilities to test wines.}
In fact, what are reasons for existence of such a space? We point out that in general 
the space $\Omega$ cannot be identified with just the collection of bottles. Alice chooses a bottle 
and her preferences are not completely determined until she makes 
the wine order.  If one likes it is possible to use the terminology that is typically used in discussions on 
foundations of quantum mechanics: ``death of reality.'' However, in this game framework this death of 
reality does not look mystically. This only means that one is not able from the very beginning 
to assign to any bottle choice and test preferences of Alice and Bob. Nothing more. Nevertheless, reality 
of wines could not be denied and choices and tests of Alice and Bob are based on this wine-reality. 

By using Gudder's theory of probability manifolds \cite{GD0} we can say that we have a probability manifold
with the atlas having two charts, one serves for the first part of game and another for the second; in Accardi's
terminology this is a non-Kolmogorovian model (he always emphasized the role of violation of Bayes' formula, 
see \cite{AC1}).

We emphasize that the choice c) of the transition probabilities implies immediately that the {\it coefficients 
of interference}  $\lambda$ are equal zero. 

\section{Action at the distance?}

One can consider Wine Game involving facelogy:  Bob  can extract some information about the origin of wine 
by observing the behavior of Alice after she has done her choice. By using the terminology of QM    one can say 
that there is ``action at the distance.'' However, even if such an action is present in the game it is not 
instantaneous! Everything happens in the complete accordance with laws of special relativity: light is reflected 
from Alice's face and Bob obtains information only when the light wave will come to his eyes.

Consideration of QL-games of the facelogy-type extends essentially the range of possible applications of our model.    
However, we do not couple directly such an action at the distance with essentially nonclassical 
probabilistic structure. The origin of nonclassicality is the impossibility to combine all possible 
preferences in a single probability space. Again by using the terminology of QM one can say that there are
two incompatible measurement settings (corresponding to two parts of Wine Game); thus we proceed in the complete 
accordance with Bohr's ideology \cite{Bohr}. 

\section{Wine Game with three players}

We now generalize Wine Game by considering three players, Alice, Bob, Cecilia. 
The first part: Alice chooses a bottle, Bob 
tests;  the second part: Bob chooses, Cecilia tests, and the third part: Cecilia chooses, Alice tests. 
We shall use
probabilities with indexes $a,b,c$ corresponding to Alice, Bob, Cecilia. For each part of the game we 
fix payment matrices. We consider symmetric game. We can write averages: 
for the first part -- $E_1^b(C) (=- E_1^a(C)),$ 
for the second part  $E_2^c(C) (=- E_2^b(C)),$ and for the third part $E_3^a(C) (= - E_3^c(C)).$ 

We assume that conditions R1)--R3) which guarantee the possibility to apply QLRA hold for all pairs 
of observables. Thus we apply QLRA to the probabilities corresponding to the pair $a,b.$ We obtain the complex
probability amplitude $\psi_{C;ab}$ which belongs to two dimensional Hilbert space which  is denoted $H_{ab}.$ Observables
$a, b$ are represented by self-adjoint operators $\hat{a}, \hat{b}$ which have bases of eigenvectors   
$\{ e_\alpha^{a;ab}\}, \{e_\beta^{b;ab}\}.$ We also apply QLRA to the probabilities corresponding to the 
pair $b,c.$ We obtain a new complex probability amplitude $\psi_{C;bc}$ which belongs to two dimensional 
Hilbert space which  is denoted $H_{bc}.$ Observables
$b, c$ are represented by self-adjoint operators $\hat{b}, \hat{c}$ which have bases of eigenvectors   
$\{e_\beta^{b;bc}\}, \{e_\gamma^{c;bc}\}.$ Finally, consider the $H_{ca}$-representation.

These representations can be identified with the aid of unitary maps:
$$
U_{ab,bc}: H_{ab} \to H_{bc},  e_\beta^{b;ab} \to e_\beta^{b;bc},
$$
and  
$$
U_{bc,ca}: H_{bc} \to H_{ca},  e_\gamma^{c;cb} \to e_\gamma^{c;ca}.
$$
The crucial point is that $U_{ab,bc}(\psi_{C;ab})=\psi_{C;bc}$ and $U_{bc,ca}(\psi_{C;bc})=\psi_{C;ca}.$
Therefore we can identify complex probability amplitudes $\psi_{C;ab}, \psi_{C;bc}, \psi_{C;ca}$ and consider
a unit vector $\psi_C$ as representing the wine collection and preferences of Alice, Bob and Cecilia.
We shall come back to this game little bit later.

We remark that this game has the structure of Gudder's probability manifold with the atlas having three charts. 

\section{Simulation of  Wine Game}

Typically quantum probabilities are imagined as rather mysterious things. Absence of the underlying 
Kolmogorv space may only support such a viewpoint. However, by using the frequency (von Mises) approach
quantum probabilities can be easily simulated. One need not use special ``quantum coins'' given by sources 
of photons or electrons. We simulate our game by using the following system of dichotomous random generators 
(taking values F and I):
$$
g_a, g_b, g^{b \vert a}(\alpha), g^{a \vert b}(\beta).
$$ 
Here $g_a$ and $g_b$ simulate choices of wine from the collection $C$ (by Alice and Bob, respectively);
the  frequencies of F and I approaches the corresponding probabilities $p_C^a(F), p_C^a(I),
p_C^b(F), p_C^b(I)$ when the number of trial goes to infinity. The generator
$g^{b \vert a}(\alpha)$ describes ability of Bob  to analyze 
wine's origin under the condition that Alice selects a bottle of the $\alpha$-origin.
 For example, the generator $g^{b \vert a}(F)$ takes the value $F$ if Bob correctly recognized 
French wine (which was chosen by Alice).  The generator $g^{a \vert b}(\beta)$ has a similar meaning.

Now, to simulate Wine Game, we just apply these generators consequently  in the right order, e.g.,
first $g_a$ and if it takes the value F, then the generator $g^{b \vert a}(F).$ That's all! We shall simulate 
probabilities and payoffs given by the two dimensional QL-model.

In all previous considerations we started with some collection of probabilities and transition probabilities 
and under the conditions R1)--R3) we were able to represent Wine Game in the two dimensional complex Hilbert
space. By applying QLRA we constructed the wave function and operators $\hat{a}, \hat{b}.$ 

We can also proceed in the opposite way. We can take two noncommutative operators in the two dimensional Hilbert
space, say $\hat{a}$ and $\hat{b},$ and a normalized vector $\psi$ in this space. Then we find 
(by using Born's rule) all probabilities which we need for Wine Game. Those probabilities will automatically 
satisfy conditions R1)--R3). Finally, we can simulate Wine Game by using the above scheme.

This strategy is especially convenient for generalizations of Wine Game to spaces of high dimension.
QLRA becomes very complicated \cite{KHC}. Reconstruction of the wave function is not so simple task. Therefore one can start
just with probabilities which are obtained from the mathematical formalism of quantum mechanics. Moreover,
the possibility to apply QLRA is restricted by a number of conditions, e.g., R1), R2). One can ignore these conditions
by starting directly with a normalized  vector $\psi.$ 

\section{Bell's inequality:  the two dimensional representation} 

We now come back to the Wine Game with three palavers, Alice, Bob, Cecilia.
We shall use the pragmatic strategy proposed at the end of the previous section. We take probabilities
and operators corresponding to a known quantum system and simulate Wine Game on the basis of these 
probabilities. We emphasize that we take probabilities given by the mathematical apparatus of quantum
mechanics and not at all a quantum physical system by itself. We introduce a game parameter $\theta \in [0, 2\pi).$
Alice is characterized by $\theta= \theta_1,$ Bob by $\theta= \theta_2,$ Cecilia by $\theta= \theta_3.$  
We take the transition probabilities
corresponding to  ``spin 1/2 system.'' For the first part of the game we have: 
$$
p^{b \vert a}(b=F \vert a=F)=p^{b \vert a}(b=I \vert a=I) =\cos^2 \frac{\theta_1 - \theta_2}{2};
$$
$$
p^{b \vert a}(b=I \vert a=F)=p^{b \vert a}(b=F \vert a=I) =\sin^2 \frac{\theta_1 - \theta_2}{2}.
$$
The transition probabilities for other parts of the game are defined in a similar way, e.g.:
$$
p^{c \vert b}(c=F \vert b=F)=p^{c \vert b}(c=I \vert b=I) =\cos^2 \frac{\theta_2 - \theta_3}{2}.
$$
Let us choose the following Darice of payoffs: 
$$
h_{FF}=h_{II}= +1, \; h_{IF}=h_{FI}= -1.
$$
Let us now suppose that Alice, Bob and Cecilia selects wine from the collection by using uniform random generators:
$p_C^a(\alpha) =p_C^b(\beta) =p_C^c(\gamma)=1/2.$
We now find the average for Bob's wins-losses in the first part of the Wine Game: 
\begin{equation}
\label{BIU0}
E_1^b \equiv E(\theta_1,\theta_2)= \cos^2 \frac{\theta_1 - \theta_2}{2} - \sin^2 \frac{\theta_1 - \theta_2}{2}=
 \cos(\theta_1 - \theta_2).
\end{equation}
In the same way we have for Cecilia:
\begin{equation}
\label{BIU0X}
E_2^c\equiv E(\theta_2,\theta_3)= \cos(\theta_2 - \theta_3)
\end{equation}
and finally for Alice: 
\begin{equation}
\label{BIU0Y}
E_3^a \equiv E(\theta_3,\theta_1)= \cos(\theta_3 - \theta_1).
\end{equation}
We set now F=+1 and I=-1, we recall that with these notations we can represent
$$
E_1^b= \rm{cov}(a,b), E_2^c= \rm{cov}(c,b), E_3^a= \rm{cov}(c,a),
$$
where covariations are taken with respect to probabilities $p^{ab}_C, p^{bc}_C, p^{ca}_C.$
We now ask: Can one construct a probability measure ${\bf P}$ and realize observables $a,b,c$ by random
variables on the corresponding Kolmogorov space in   such  a way that
\begin{equation}
\label{BIU}
{\bf P}(a=\alpha, b=\beta)=   p^{ab}_C(a=\alpha, b=\beta), \;
{\bf P}( b=\beta, c=\gamma)=   p^{bc}_C(b=\beta,c=\gamma), \; 
\end{equation}
\begin{equation}
\label{BIU1}
{\bf P}(c=\gamma, a=\alpha)=   p^{ca}_C(c=\gamma, a=\alpha)?
\end{equation}
The answer is negative.  If representations (\ref{BIU}), (\ref{BIU1}) can be constructed, then one can prove 
Bell's inequality \cite{BEL}, see \cite{KH3} for details:
\begin{equation}
\label{BBB} \vert \rm{cov}(a,b) - \rm{cov}(b,c)\vert \leq 1 - \rm{cov}(c,a). 
\end{equation}
But it is known that Bell's inequality is violated for 
covariations given by (\ref{BIU0})--(\ref{BIU0}) for some choices of parameters (one could also apply Vorobjev's theorem
\cite{VR}). 

Thus if there is a classical probabilistic model behind Wine Game (for some set of probabilities), then
averages of payments satisfy the following Bell's inequality:
\begin{equation}
\label{BBB1} \vert E_1^b - E_2^c\vert \leq 1 - E_3^a. 
\end{equation}
Even intuitively it is clear that there are no reasons to assume that this inequality should holds for any 
set of probabilities.

The expression in the left-hand side of the Bell's inequality is equal to the 
average of the total win-loss of Bob in the game (i.e., in the two series of games -- with Alice and Cecilia, 
in the first Bob tests wine and the second Cecilia does this): $E^b =E_1^b + E_2^b= E_1^b - E_2^c=
\cos(\theta_1 - \theta_2) - \cos(\theta_2 - \theta_3).$

\section{Multidimensional games}

Wine Game can be generalized to the Hilbert space $H$ of an arbitrary dimension. 
The only difference is that now the collection $C$  contains wines from $n$ countries,
which are labeled by $i=1,...,n.$ 

Let us consider two self-adjoint operators $\hat{a}$ and $\hat{b}$ and corresponding orthonormal bases
of eigenvectors $\{ e_i^a \}_{i=1}^n$ and $\{ e_j^b\}_{j=1}^n.$ We remark that in general
we do not suppose the validity of the condition R2). In principle, operators could even commute. Of course,
QLRA would not work in such a case. But our task is not reconstruct probabilities from the game, 
but only to simulate the game.   

We also take a normalized vector $\psi \in H.$
This vector $\psi$ describes collections of wines created by Alice and Bob as well as their experiences 
of testing of wines. Actions of Alice and Bob are now labeled by $i=1,...,n.$ The tree of this extensive form game
have $n$ nodes leaving this vertex. The Bob's average is given by 
$$
E^b=\sum_{i,j=1}^n h_{ji;1}^b \; \vert \langle \psi_C, e_j^a \rangle\vert^2 \;\vert \langle e_i^b, e_j^a \rangle \vert^2 
+ \sum_{i,j=1}^n  h_{ij;2}^b \; \vert \langle \psi_C, e_i^b \rangle\vert^2 \;\vert \langle e_i^b, e_j^a \rangle \vert^2.
$$ 

To find conditions when the game has no underlying classical probability space, one can apply Vorobjev's theorem
which was proved for multi-valued random variables.


\begin{thebibliography}{99}

\bibitem{Bohr}  N. Bohr, {\it The philosophical writings of Niels
Bohr}, 3 vols. (Woodbridge, Conn., Ox Bow Press, 1987).

\bibitem{PL1} A. Plotnitsky, {\it The knowable and unknowable} 
(Univ. Michigan Press, 2002).

\bibitem{PL2} A. Plotnitsky, {\it Found. Phys.}, {\bf 33}, 1649 (2003).

\bibitem{AC1}  L. Accardi, Phys. Rep. {\bf 77}, 169(1981);
``The probabilistic roots of the quantum mechanical paradoxes,''
in {\it The wave--particle dualism:  A tribute to Louis de Broglie on his 90th
Birthday}, S. Diner, D. Fargue, G. Lochak, and F. Selleri, eds.
(D. Reidel Publ. Company, Dordrecht,  1984), pp. 47-55.

\bibitem{AC2} L. Accardi, {\it Urne e Camaleoni: Dialogo sulla realta,
le leggi del caso e la teoria quantistica} (Il Saggiatore, Rome, 1997); ``Locality
and Bell's inequality'', Q. Prob. White Noise Anal. {\bf  13},1 (2001).

\bibitem{AC3} L. Accardi, A. Fedullo, Lettere al Nuovo Cimento {\bf 34,} 161-172  (1982).

\bibitem{AC4} L. Accardi, Il Nuovo Cimento B {\bf 110}, 685 (1995).

\bibitem{BL1}  L. E. Ballentine,  Rev. Mod. Phys. {\bf 42},  358 (1970).

\bibitem{BL2} L. E. Ballentine, {\it Quantum mechanics} (Englewood Cliffs,
New Jersey, 1989).  

\bibitem{BL3} L. E. Ballentine, ``Interpretations of probability and quantum theory,''
Q. Prob. White Noise Anal. {\bf  13},  71 (2001).

\bibitem{BL4}  L. E. Ballentine, {\it Quantum mechanics} (WSP,  Singapore, 1998).

\bibitem{DM1} W. M. De Muynck, ``Interpretations of quantum mechanics,
and interpretations of violations of Bell's inequality'',
Q. Prob. White Noise Anal. {\bf  13},  95 (2001).

\bibitem{DM2} 8. W. M. De Muynck, {\it Foundations of quantum mechanics, an empiricists approach} (Kluwer, Dordrecht,
2002).

\bibitem{GD1} S. P. Gudder, Trans. AMS {\bf 119},  428 (1965).


\bibitem{GD0} S. P. Gudder, {\it Axiomatic quantum mechanics and generalized probability theory}
(Academic Press, New York, 1970).

\bibitem{GD2} S. P. Gudder, ``An approach to quantum probability,''
Quantum Prob. White Noise Anal. {\bf  13}, 147 (2001).

\bibitem{LA} A. Land\'e, {\it Foundations of quantum theory} (Yale Univ. Press, 1955).

\bibitem{LA1} A. Land\'e, {\it New foundations of quantum mechanics} (Cambridge Univ. Press, Cambridge, 1968).

\bibitem{MC1}  G. W. Mackey, {\it Mathematical foundations of quantum mechanics}
(W. A. Benjamin INc, New York, 1963).

\bibitem{FPP} A. Yu. Khrennikov, editor,  {\it Foundations of
Probability and Physics,} Ser.  Quantum Probability and White Noise
Analysis {\bf 13}, WSP, Singapore, 2001.

\bibitem{FPP1} A. Yu. Khrennikov, editor, {\it Quantum
Theory: Reconsideration of Foundations,} Ser. Math. Modeling  {\bf
2}, V\"axj\"o Univ. Press, V\"axj\"o, 2002.

\bibitem{FPP3} A. Yu. Khrennikov, editor,  {\it
Foundations of Probability and Physics-2,} Ser. Math. Modeling {\bf
5}, V\"axj\"o Univ. Press,  V\"axj\"o, 2003.

\bibitem{P1} R. Penrose, {\it The emperor's new mind} (Oxford Univ. Press, New-York, 1989).

\bibitem{P2} R. Penrose, {\it Shadows of the mind} (Oxford Univ. Press, Oxford, 1994).

\bibitem{KH1} A. Yu. Khrennikov, {\it Information dynamics in cognitive,
psychological, social,  and anomalous phenomena} (Kluwer, Dordreht,
2004).

\bibitem{KH2} A. Yu. Khrennikov,  {\it BioSystems}, {\bf 84}, 225-241 (2006).

\bibitem{VN1} J. von Neumann, O. Morgenstern, {\it Theory of Games and Economic Behaviour}
(Princeton Univ. Press, Princeton, 1953).

\bibitem{OW} G. Owen,  {\it Game theory}  (W.B.Saunders Company, Philadelphia-London-Toronto, 1968).


\bibitem{D} P. A. M. Dirac, {\it  The Principles of Quantum Mechanics}
(Oxford Univ. Press, Oxford,  1930).


\bibitem{VN} J. von Neumann, {\it Mathematical foundations
of quantum mechanics} (Princeton Univ. Press, Princeton, N.J.,   1955).


\bibitem{KOL}  A. N. Kolmogoroff, {\it Grundbegriffe der Wahrscheinlichkeitsrechnung}
(Springer Verlag, Berlin, 1933); reprinted:
{\it Foundations of the Probability Theory}
(Chelsea Publ. Comp., New York, 1956).

\bibitem{VR} N. N. Vorob'ev,  {\it Theory Prob. and its Appl.} {\bf 7}, 147-162 (1962).

\bibitem{BEL} J. S. Bell, {\it Speakable and unspeakable in quantum
mechanics} (Cambridge Univ. Press, Cambridge, 1987).

\bibitem{EF} R. Cressman, {\it Evolutionary Dynamics and Extensive Form Games.}
(MIT Press, Cambridge, MA, 2004).

\bibitem{OS} M. J. Osborne, A. Rubinstein, {\it A course in game theory} (MIT Press, Cambridge, MA,
1994).

\bibitem{KHC} A. Yu. Khrennikov, {\it J. Phys.A: Math. Gen.} {\bf 34}, 9965 (2001);
{\it J. Math. Phys.}{\bf 44}, 2471 (2003); {\it Phys. Lett. A} {\bf 316}, 279 (2003);
{\it Annalen  der Physik} {\bf 12}, 575 (2003); {\it Foundations of Physics}  {\bf 35}, 
1655 - 1693 (2005); {\it J. Math. Phys.} {\bf 45},  902-921 (2004).

\bibitem{MI} R.  von Mises, {\it The mathematical theory of probability and
 statistics} (Academic, London,  1964).

\bibitem{KH3} A. Yu. Khrennikov, {\it Interpretations of probability} (VSP Int. Sc. Publ.,
Utrecht, 1999).

\bibitem{GR1} A. A. Grib, G. N. Parfionov, ``Can the game be quantum?'' Notes of Sc. Sem. 
          Petersburg's Branch Math. Inst. Russian Acad. Sc., {\bf 291}, 1-24 (2002).

\bibitem{GR0} A. A. Grib, A. Yu. Khrennikov, K. Starkov, ``Probability amplitude in quantum-like games,''
in {\it Quantum Theory: Reconsideration
of Foundations} (Ser. Math. Modelling, V\"axj\"o Univ. Press, V\"axj\"o, 2004), {\bf 10},
pp. 703-722.

\bibitem{GR0X} A. A. Grib, A. Yu. Khrennikov,  G. N. Parfionov, K. A. Starkov,
 ``Distributivity breaking and macroscopic quantum games,'' in
{\it Foundations of probability and physics---3}  (AIP Conf.
Proc.,  Amer. Inst. Phys., Melville, NY, 2005), {\bf 750}, pp. 108-113.

\bibitem{GR3} A. Grib, A. Khrennikov, G. Parfionov, and K. Starkov, {\it J. Phys. A.: Math. Gen.},
{\bf 39}, 8461-8475 (2006).

\bibitem{GR4} A. A. Grib, R. R. Zapatrin.  Int. J. Theor. Phys.  {\bf 29}(2), 113-123 (1990).

\bibitem{GR5} A. A. Grib, W. A. Rodrigues, {\it Nonlocality in Quantum Physics} (Kluwer Academic Publ./
Plenum Publishers, New York-Boston-Dordrecht-London-Moscow, 1999).


\bibitem{Ek} A. K. Ekert, Phys. Rev. Lett. {\bf 67}, 661 (1999).

\bibitem{ES} B. d'Espagnat, {\it Veiled Reality. An anlysis of present-day
quantum mechanical concepts} (Addison-Wesley, 1995).

\bibitem{Sh} A. Shimony, {\it Search for a naturalistic world view} (Cambridge Univ. Press, Cambridge, 1993).


\end{thebibliography}
\end{document}